\documentclass[12pt,preprint]{aastex}

\newcommand{\myemail}{maggie@physics.mcgill.ca}
\newcommand{\psr}{PSR~B1509$-$58}

\newcommand{\tempo}{{\tt{TEMPO}}}
\newcommand{\nudotdotdot}{{\ifmmode\stackrel{\bf \,...}{\textstyle \nu}\else$\stackrel{\,\...}{textstyle \nu}$\fi}}

\usepackage{graphicx}

\shorttitle{Long-term Timing of \psr}
\shortauthors{Livingstone et al.}

\begin{document}
\title{21 years of Timing \psr.}

\author{Margaret A.~Livingstone \altaffilmark{1},
Victoria M.~Kaspi, Fotis P.~Gavriil}
\affil{Department of Physics, Rutherford Physics Building, 
McGill University, 3600 University Street, Montreal, Quebec,
H3A 2T8, Canada}

\and
\author{Richard N.~Manchester}
\affil{Australia National Telescope Facility, CSIRO, P.O. Box 76,
Epping, NSW 1710, Australia}

\altaffiltext{1}{\myemail}

\clearpage

\begin{abstract}
We present an updated timing solution for the  young, energetic pulsar \psr\
based on 21.3 years of radio timing data and 7.6 years of X-ray
timing data. No glitches have occurred in this time span, in contrast
to other well-studied young pulsars, which show frequent glitches. We report a 
measurement of the third frequency derivative of 
$\nudotdotdot=(-1.28 \pm 0.21) \times 10^{-31}\,$s$^{-4}$. This value is
1.65 standard deviations from, i.e. consistent with, 
that predicted by the simple constant magnetic dipole
model of pulsar spin-down. We measured the braking index to be $n=2.839 \pm 0.003$
and show that it varies by 1.5\% over 21.3\,yr due to contamination from
timing noise. Results of a low-resolution power spectral analysis of the significant noise
apparent in the data yield a spectral index 
of $\alpha = -4.6 \pm 1.0$ for the red noise component.
\end{abstract}


\section{Introduction}
\label{sec:intro}
Radio pulsars are powered by their
rotational kinetic energy: as they emit electromagnetic radiation, their rotation
rates decrease. Thus, measuring the rotational evolution of pulsars is a probe 
of the physics underlying these fascinating objects. Young, energetic pulsars, such 
as \psr\  are especially interesting due to their large spin-down luminosities and rapid
spin-down rates. High rates of spin-down allow for the measurement 
of higher-order frequency
derivatives, providing a deeper probe into the physics of spin-down. This 
slow-down can be described by
\begin{equation}
\dot{\nu} = -K{\nu}^n      ,
\end{equation}
where ${\nu} \equiv 1/P$ is the pulse frequency,
$\dot{\nu}$ is its derivative and $n$ is the braking index. For braking by
magnetic-dipole radiation in a vacuum, $K$ is related to the dipole magnetic
moment of the pulsar and $n=3$ \citep{mt77}. Differentiating (1) shows that $n$ is given by
\begin{equation}
n=\frac{ {\nu}{\ddot{\nu}}}{\dot{\nu}^2}  .
\end{equation}
The braking index is only measurable
for the youngest pulsars and values not entirely dominated by noise processes 
have been measured for only 5 objects. In all cases, the measured values are
less than the canonical value of $n=3$ \citep{lps93,kms+94,lpgc96,dnb99,ckl+00}.
Implications of observed braking indices less than 3 are still being 
discussed \citep[see for example, ][]{br88}
and include magnetic field growth
or alignment with the spin axis, effects due to higher-order magnetic
moments and the interaction of the pulsar with its relativistic particle
wind. 

Most pulsars do not have
measurable braking indices due to a combination of contamination from low-frequency noise
processes and the relatively slow spin-down of middle-aged pulsars. Because
of the form of
the spin-down power law given in equation 1, successive
frequency derivatives go approximately as powers of the first frequency
derivative. Thus most pulsars, which have values of $\dot{\nu}$ of
$\sim 10^{-15}$\,s$^{-2}$, should exhibit $\ddot{\nu}$ on the order of
$\sim 10^{-30}$\,s$^{-3}$, which would require hundreds of years of observation to
measure. Many younger pulsars that could conceivably have measurable
$\ddot{\nu}$ exhibit significant timing noise which masks
the deterministic value of $\ddot{\nu}$ due to dipole spin-down, resulting in measured
values orders of magnitude larger than predicted, often with the wrong sign.
Furthermore, these values are not stable and consequently do not accurately
predict pulse arrival times.
\par
An important check on the validity of the spin-down law is the measurement
of deterministic higher-order frequency derivatives. An expression
for $\nudotdotdot$ is
given by taking an additional derivative of equation (1),
\begin{equation}
\nudotdotdot\ = \frac{n(2n-1){\dot{\nu}^3}}{\nu^2}. 
\end{equation}
We define the second braking index to be $m_0 \equiv n(2n-1)$. If equation 1
accurately characterizes the spin-down of the pulsar, then $m_0=m$, where
\begin{equation}
m = \frac{{\nu}^2 \nudotdotdot}{\dot{\nu}^3}.
\end{equation}
A measurement of $m$ provides important insight into the spin-down of
young pulsars and $m \neq m_0$ implies that the characteristic 
age is an under- or over- estimate of the true age, 
as well as having implications for a changing magnetic field \citep{bla94}. 
Evidently, the measurement of higher-order frequency derivatives will only 
be possible in the youngest pulsars that have large spin-down rates
not dominated by noise processes. The Crab pulsar has a measured value
of $\nudotdotdot = (-6.45\pm0.02)\times 10^{-31}\,$s$^{-4}$ \citep{lps93}, in
agreement with the prediction from the spin-down law, in spite of probable
contamination from timing noise and frequent glitches. 

\psr\ has 
never been observed to glitch, hence is an excellent candidate to have a
measurable $\nudotdotdot$, as well as a very accurate measurement
of $n$. \citet{kms+94}  measured a value for
$\nudotdotdot = (-1.02 \pm 0.25) \times 10^{-31}\,$s$^{-4}$, based on 11\,years
of radio timing data, implying
$m=14.5 \pm 3.6$, in agreement with the predicted value $m_0=13.26\pm 0.03$. 
The uncertainty in $\nudotdotdot$ and thus $m$ was known to be larger than
the formal uncertainty due to contamination from significant timing noise in
the data. To account, albeit not rigorously, for this effect, the quoted
uncertainty is three times the formal uncertainty. 
\par
In this paper, we extend the timing analysis performed by
\citet{kms+94} using an additional 10.3\,yr of radio data as well as
7.6\,yr of X-ray timing data to increase 
the precision with which the frequency evolution of \psr\ can be measured.
This is accomplished by performing a phase-coherent analysis of the the data
as well as a partially phase-coherent analysis that is less sensitive to the
timing noise superimposed on the deterministic spin-down. In addition, we
present a low-resolution spectral analysis of the timing noise and discuss 
possible physical causes of the noise process. 

\section{Observations}
\label{sec:obs}
\subsection{Radio Timing data}
\label{sec:radiotiming}
In our 21-yr data span, observations were made with a variety of different
data acquisition instruments, at two different radio telescopes.
Observations of \psr\ were conducted using the Molonglo Observatory Synthesis
Telescope (MOST) at an observing frequency of 843\,MHz from 1982 June 24
through 1988 June 23. Our analysis includes a total of 177 MOST pulse
arrival times. For full details of these observations, see \citet{mdn85}.
All data used after 1990 were obtained using the Parkes 64-m telescope.
Observations beginning 1990 March 15 and continuing at roughly monthly
intervals until 21 January 1994 were described in \citet{kms+94}. The same
observing system, consisting of a 64-channel analog filterbank spanning
320\,MHz of bandwidth at a central frequency near 1400\,MHz, was used until
1997 January. Our analysis includes a total of 127 pulse arrival times
obtained with the 64-channel filterbank system at Parkes. From 1997 May
through 2003 October, filterbank observations were made near the same
central frequency but using a 96-channel filterbank spanning 288\,MHz of
bandwidth. Details of this observing system, the central beam of the Parkes
Multibeam receiver, can be found in \citet{mlc+01}. A total of 43 pulse
arrival times were obtained with this system. Interspersed with the
filterbank arrival times are data obtained also at Parkes, but using the
Caltech correlator system \citep{nav94, sbm+97}. Typical Parkes integration
times were $\sim$3000\,s for all systems. In spite of the variety
of observing systems used, the Parkes data generally are of uniform quality,
with typical folded pulse signal-to-noise ratios of $\sim$20. 
\par
The Parkes
data set also includes a handful of observations at a significantly higher
radio frequency. These data, obtained primarily at four epochs, were used to
determine the dispersion measure (DM). Two observations at 2480\,MHz on 1990 March 16
were used with observations at 1400\,MHz on 1990 March 15 and 1990 March 17 
to determine the DM for the phase-coherent timing analysis and the DM
remained constant within uncertainties throughout the entire data set (see 
Section \ref{sec:dm}). Folded pulse
profiles were cross-correlated with a template
to yield topocentric arrival times for each observation. Topocentric arrival
times were fitted to a timing model (see Section 3.1) using the \tempo\ 
\footnote{http://www.atnf.csiro.au/research/pulsar/tempo/} software package. 

\subsection{X-ray timing data}
\label{sec:xray}
The results presented in the X-ray analysis 
were obtained with public data from the Proportional
Counter Array \citep[PCA;][]{jsg+96} on board the Rossi X-ray Timing
Explorer (\textit{RXTE}). The PCA consists of an array of five collimated
xenon/methane multi-anode proportional counter units (PCUs) operating
in the 2\,--\,60~keV range, with a total effective area of approximately
$\rm{6500\,cm^2}$ and a field of view of $\rm{\sim 1^o}$\,FWHM.
We used 7.6\,yr of archival \textit{RXTE}\ observations  collected in
the ``GoodXenonwithPropane'' mode, which records the arrival time
(with 1-$\mu$s resolution) and energy (256 channel-resolution) of every
unrejected xenon event as well as all the propane layer events. 
We used all xenon layers of each PCU in the 2--32\,keV
range because of the relative hardness of the source. 
The observations were reduced using software developed at MIT for handling
raw spacecraft telemetry packet data. Data from the different PCUs 
were merged and binned at 1/1024\,ms resolution. The data were then 
reduced to barycentric dynamical time (TDB) at the solar system barycenter 
using the known position from radio interferometry
(see Table 1) and JPL DE200 solar system ephemeris. Each 
time series was folded with 64 phase bins using the contemporaneous radio timing ephemeris. 
Resulting pulse profiles were cross-correlated in the Fourier domain with 
a high signal-to-noise-ratio template created by adding phase-aligned 
profiles from previous observations. We implemented a Fourier domain 
filter by using only the first 6 harmonics in the cross-correlation due to
the relatively sinusoidal pulse profile. The
cross-correlation produces an average time of arrival (TOA) for each 
observation. 

\section{Results}
\label{sec:timing}
\subsection{Phase coherent timing analysis}
\label{sec:coherent}
The stable rotation of pulsars allows (in the absence of glitches) 
a phase-coherent analysis of the
timing data, that is, each turn of the pulsar is accounted for. This is 
accomplished by fitting the TOAs to a Taylor expansion of pulse
phase, $\phi$. Pulse phase at any time $t$ can be expressed as
\begin{equation}
\phi(t)=\phi(t_0) + \nu_0(t-t_0) + \frac{1}{2}{\dot{\nu}_0}(t-t_0)^2 +
        \frac{1}{6}{\ddot{\nu}_0}(t-t_0)^3 
	+ \frac{1}{24}{\nudotdotdot_0}(t-t_0)^4 {\ldots} , 
\end{equation}
where the subscript `0' denotes a parameter evaluated at the reference epoch
$t_0$. Both the radio and X-ray TOAs were fit to the above polynomial using
the software package {\tempo}.  Figure 1 shows timing 
residuals from a phase-coherent timing analysis with {\tempo} with 
radio timing data shown as dots
and X-ray data shown as crosses. The position of the pulsar was held fixed
at the position determined by radio interferometry 
\citep[Table 1;][]{gbm+99}. A constant phase
offset was fitted between the radio and 
X-ray data to allow for the differences between the radio and X-ray
pulse profiles. The top panel shows residuals with $\nu, 
\dot{\nu}, 
\ddot{\nu}$ and $\nudotdotdot$ fitted out. The middle panel shows the fourth
frequency derivative also fitted; the bottom panel shows the residuals with
the fifth frequency derivative fitted. The large concentration of power in
the fifth frequency derivative is not predicted by theory, but is presumably
due to timing noise. The absence of any sudden discontinuties visible in the 
residuals (Figure 1) indicates that no glitches have occured in the 21.3\,yr of
observations.
\par
The usual method of determining timing parameters
involves fitting many higher-order derivatives to `pre-whiten' the residuals
\citep[e.g.][]{ktr94}. However, we found that the value of 
$\nudotdotdot$ changes with each higher-order derivative fit, without 
converging to a single value, and without
entirely `whitening' the residuals. This is due to two effects:
contamination from timing noise and covariance in the fitted
parameters. The value of $\nudotdotdot$ is ambiguous from this type of
analysis. A similar effect is seen with $\ddot{\nu}$, though to a much smaller
level (i.e. within formal uncertainties) and is thus unimportant.
\par
Nevertheless, it is clear from 
the timing residuals that there is a very
significant signal from $\nudotdotdot$ and it is unlikely that it is entirely
due to timing noise. Hints that a frequency derivative may be dominated by a
noise process are if the
value is several orders of magnitude larger than the value predicted from
theory, if it is of the wrong sign, and if it does not correctly predict
pulse arrival times. None of these are true for \psr, indicating that though
there is certainly some contamination from timing noise, it is likely that
the noise component does not dominate. Thus to 
determine $\nudotdotdot$, we used a partially phase-coherent timing analysis 
as described in the next section. Spin parameters determined by the 
phase-coherent timing analysis are given in Table 1. 

\subsection{Partially phase-coherent timing analysis}
\label{sec:incoherent}
To determine $\nudotdotdot$ unambiguously, we employed a partially 
phase-coherent timing analysis. This
method can also be used when a fully coherent solution 
is impossible due to glitches or large gaps in the data. We 
split the 21.3-yr of data into subsets of approximately
2\,yr and fitted each phase-coherently. We ensured that the residuals
were white for each interval and that the covariances 
between the fitted parameters 
($\nu$, $\dot{\nu}$ and $\ddot{\nu}$) were less than 0.9. The results are
shown in Figure 2. The
slope of the line, determined by a weighted least squares fit, 
is $\nudotdotdot$ and has a value of $(-1.28\pm 0.21)\times
10^{-31}\,$s$^{-4}$.  This partially coherent method heavily samples
the data points at the ends of each interval, while weakly sampling the
middle, so we also considered intervals
that better sampled the mid-point as much as possible.  We 
used these latter data points only to confirm the measurement of
$\nudotdotdot$, but have
not included them in the quoted value because it was impossible to
optimize these points in the same manner as the first set.  The 
uncertainty was obtained from a bootstrap analysis, as we
suspected that 
the formal uncertainty underestimated the true uncertainty
due to contamination from timing noise. The bootstrap is a
robust method of determining errors when a small number of sample
points is available, as in this case \citep{efr79}.  

In addition to measuring $\ddot{\nu}$, we measured the
braking index for each subset.  The results are shown in Figure 3. The 
weighted average value is $2.839\pm0.003$. A weighted linear least-squares
fit showed that $\dot{n}$ is consistent with zero, i.e. that there is
no underlying constant trend in the data. There is, however, significant variation
from the average value; the reduced $\chi^2$ is 15 for 9 degrees of
freedom. This value of $n$ is consistent with both the measurement from the fully
coherent timing analysis (see Table 1) and that measured by \citet{kms+94}. 

\subsection{Timing noise}
\label{sec:noise}

Despite the renowned steady rotation of pulsars, there is timing
noise apparent in most pulsar timing data. A low-frequency timing noise 
process can be seen in the residuals of \psr\ (Figure 1). These processes are
typified by long-term polynomial-like trends in timing residuals after
all determinstic spin-down effects have been removed. Timing noise has been
shown to be correlated with $\dot{P}$, thus tends to be most apparent in young
pulsars \citep{cd85, antt94}. The low-frequency nature of the noise 
process (giving rise to the adjective `red'), complicates both its removal 
and understanding.

One common way of characterizing timing noise is with the $\Delta_8$
parameter, given by $\Delta_8 =$log$(\frac{1}{6\nu}|{\ddot{\nu}}|t^3)$
\citep{antt94}. This parameter estimates how much the timing noise in 
the $\ddot{\nu}$ term contributes to the cumulative phase of the 
pulsar, assuming that the measurement of $\ddot{\nu}$ {\it{is 
dominated by timing noise}}. One must be careful when calculating 
the $\Delta_8$ parameter for pulsars that have 
deterministic values of $\ddot{\nu}$, since these are no longer dominated by a
noise process. Instead, the contribution to $\ddot{\nu}$ from timing noise
must be found. We estimated the noise in this term using the partially
coherent analysis explained in the previous section, by calculating the root
mean square of the scatter about the deterministic trend. However, we found
the uncertainty in the rms value to be larger than the rms value
itself, so the value of $\Delta_8$ is consistent with zero. This is not
inconsistent with the prediction made by \citet{antt94}, given the large
amount of scatter in their data. 

For a
general review of timing noise in pulsars, see papers by
\citet{cor80}, \citet{ch80} and \citet{cg81}. The physical causes 
of timing noise in pulsars
are poorly understood; possibilities range from changes in moment of inertia
due to random pinning and unpinning of superfluid vortices in
the core of the neutron star \citep{cg81} to torques acting on the crust due
to interactions between the pulsar and its magnetosphere
\citep{che87a}, or that isolated pulsars may be experiencing 
free precession \citep{sls00}. 

Another possible cause of apparent timing noise in some objects is the 
presence of one or more planet-mass objects orbiting the neutron star. If these
planets are of sufficiently low mass,
their presence may appear to be `timing noise' for many years. In fact,
this possibility has been suggested by \citet{arn02,arn04}
for \psr. He performed a fit consisting of five orbits of 
$\sim$solar mass planets to the same {\textit{RXTE}} timing
residuals that we have used in our analysis. To determine if these orbits
were stable, we performed a periodogram analysis on both the X-ray
timing residuals and the entire data set. We found roughly
similar periodicities in the 7.6-yr X-ray data set, but these
periodicities are not seen in the entire 21.3-yr data set. This
indicates that the `periodicities' are not stable in time and thus likely do
not represent orbits of planets around \psr.

\subsection{Dispersion Measure Variations}
\label{sec:dm}
One possible contributing factor to the observed timing noise is changes in
the DM of the pulsar. Variations in DM have
been measured for several pulsars, notably for the Crab and Vela
pulsars \citep{bhvf93, hmm+77}. It
has been postulated that the short-term DM variations observed
in the Crab are due to wisps passing through the line of sight. A 
typical variation in DM for the Crab pulsar is
$\Delta$DM=0.02\,pc~cm$^{-3}$ with a rise time of $\sim40$\,days and decay 
time of $\sim500$\,days, although more recently, a  DM variation of 
$\Delta$DM=0.15\,pc~cm$^{-3}$ was reported \citep{wbl01}. This caused a
change in arrival time of $\sim 1\,$ms, at an analysis frequency of 327\,MHz. A
similar change in DM in \psr\ would result in a change in arrival time of
$\sim 0.32$\,ms at 1400\,MHz, undetectable in our data.  

The DM for \psr\ was determined at four epochs where 
sufficient multifrequency data were obtained as shown in Table 2. Each 
value was determined by fitting only for $\nu$ and DM using 
only the multifrequency data occuring over a short time span (see Table 2). The 
DM was held fixed for the remainder of the timing analysis 
at the value obtained for the 1990 epoch. Unfortunately,
any changes in the DM that occured at epochs other than 
those with multifrequency data remain unmodelled and will thus contribute 
to the noise in the data, since the time of arrival of a pulse changes 
with DM. 

A weighted least-squares fit was performed on the four DM measurements and the best-fit 
slope gives $\Delta$DM= $(0.42\pm0.19)$\,pc~cm$^{-3}$\,yr$^{-1}$. 
Data spanning these four epochs were also fitted phase-coherently using
\tempo\ resulting in $\Delta$DM = $(0.76\pm0.25)$\,pc~cm$^{-3}$\,yr$^{-1}$.
Although this second measurement appears marginally significant, the small
variations in observing frequency throughout the data allow timing noise to
be absorbed into the $\Delta$DM measurement, thus underestimating the
uncertainties. From Table 2, we can see that the largest single change in DM 
(occuring between 1990 and 1996), $\sim 4$\,pc~cm$^{-3}$, would cause
a delay in arrival time of $\sim 8.5$\,ms at 1400\,MHz. 
There is no evidence, however, of a sudden change in DM, and a
gradual change in DM of this magnitude would be difficult to detect due to
the instrinsic scatter in the arrival times of the pulses.

\par
Another determination of DM is obtained for the time period starting in
1996, when \textit{RXTE} began observing the source. X-rays are unaffected by
changes in DM, thus by comparing the X-ray and radio residuals, we can determine
if changes in DM have occurred. We fit
the X-ray data and all contemporaneous radio data with \tempo, fitting out
many higher-order derivatives to `pre-whiten' the residuals. Though it is
impossible to fit for DM directly with \tempo\ between radio and X-ray data, we can
compare the residuals. Due to the uncertainty between the radio and X-ray 
pulse profiles, we are
required to fit a constant offset between the two data sets. Thus we are
insensitive to any constant change in DM between this and an earlier 
epoch. The timing residuals show no evidence 
for a systematic change in DM over 7.6\,yr. However, the
scatter remaining in the data is larger than expected given the TOA
uncertainties. This scatter may be, in part, due to changes in DM on
timescales of $\sim$1-2\,months, appearing random in nature in our sparsely
sampled data set. The paucity of multifrequency data results in the
inability to discern between short-term DM variations and a random noise
process that is independent of observing frequency. We can rule out changes
in DM that would change the arrival time of the pulses by more than the
scatter in the data, $\sim$10\,ms. Thus an observing frequency of 1400\,MHz, 
DM variations larger than 4.7\,pc~cm$^{-3}$ are ruled out.

\subsection{Power Spectrum Analysis}
\label{sec:spectrum}
Due to the low-frequency nature of the noise process active in most pulsars,
as well as the uneven sampling of our data, a traditional Fourier
spectral analysis is rendered ineffective as an estimator of the power spectrum
 \citep{db82}. These problems can be partially 
circumvented by using a method of extracting
power spectra based on orthogonal polynomials estimators developed by 
\citet{gro75a} and \citet{dee84}. This method provides a
low-resolution but reasonable estimate of the power contained in timing
noise over a range of frequencies. To extract a power spectrum from the 
noise component of the timing data, the deterministic spin-down 
terms (in this case, $\phi$, $\nu$, 
$\dot{\nu}$, $\ddot{\nu}$ and $\nudotdotdot$) are fit out with \tempo. The
next order term contains a large contribution from red noise (Figure 1), while
higher-order terms will contain declining contributions from red noise and a
white noise contribution attributable to measurement errors. 

First, it is necessary to construct orthonormal polynomials $p_j , j=0, 1, .., N, $
such that
\begin{equation}
\sum_i ~p_j(t_i)\,p_k(t_i) = \delta_{jk} ,
\end{equation}
where each $t_i$ represents one TOA. Due to the nature of complete sets of orthogonal polynomials, they can,
given the proper coefficients, 
describe any function. In this case, they are used to describe the timing
residuals:
\begin{equation}
R(t) = \sum_j~c_j\,p_j(t),
\end{equation}
where R(t) are the residuals, and the coefficients are fitted for and described by:
\begin{equation}
c_j = \sum_iR(t_i)\,p_j(t_i).
\end{equation}

Since ${\phi}$, ${\nu}, {\dot{\nu}}, {\ddot{\nu}}$ 
and $\nudotdotdot$, have
been already been fitted, they are completely
covariant with $c_0, c_1, c_2, c_3$ and $c_4$ and thus contribute no
additional information about the noise process. Therefore $c_5$ is
the lowest order 
coefficient that contains useful information about the timing noise. An
estimate of the power contained at the lowest frequencies, that is $f=1/T$,
where $T$ is the length of the data span, is given by $S_m =
|{{c_5}}|^2$ \citep{gro75a}. It is
possible to obtain further estimates of the power at higher frequencies by
splitting the data into $m=2, 4, 8{\ldots}$ sections, which give estimates
of the power contained at $f=m/T$. For each value of $m$, there
are $m$ estimates of $S_m$, which are averaged to give the mean power estimate.
Each average of $S_m$ is distributed as a ${\chi}^2$
distribution with $m$ degrees of freedom. The averages are then corrected
by a factor to account for the difference between the median and the mean of a ${\chi}^2$
distribution \citep{ktr94}. The results of this analysis are plotted in
Figure 4. The highest frequency term is certainly due to measurement errors,
and is of the same approximate power level as the previous three terms.
Thus, these terms are not included in the fit of the power spectrum. The slope
obtained by fitting the three lowest frequency terms is $\alpha = -4.6 \pm
1.8$. However, fitting the fourth term, (whose uncertainty places it just above
the white noise level) gives the same spectral index while 
better constraining the uncertainties, and implies a  spectral index of 
$\alpha = -4.6 \pm 1.0$. 

\section{Discussion}
\label{sec:discussion}
\subsection{Timing Analysis}
\label{sec:timing_discussion}
The measurement of the third frequency derivative, $\nudotdotdot = (-1.28
\pm 0.21) \times 10^{-31}$s$^{-4}$ is 1.65 standard deviations from 
the value predicted by the spin-down law. Hence the observed value and the 
model prediction are consistent. Typically, uncertainties on $\nudotdotdot$ 
fall as $t^{-4}$, though
the prominence of timing noise in this case will undoubtably lengthen
the amount of time required. Based on the bootstrap analysis used to
determine uncertainties, another $\sim 10$\,yr of observations are required
to reduce uncertainties by a factor of $\sim 2$. Assuming that we have 
measured a stable value
of $\nudotdotdot$ and that the uncertainties will be reduced by a
factor of 2, then the simple spin-down law derived assuming 
a constant magnetic field may not sufficiently describe the spin-down of
\psr. 

\par
The consequences of a 
measured $\nudotdotdot$ different from the predicted value are most clear
when comparing the measured and predicted values of the second 
braking index. The measured value of 
$\nudotdotdot$ implies a second braking index $m= 18.3 \pm 2.9$, larger than
$m_0=13.26\pm 0.03$, but not significantly so given the uncertainties. \citet{bla94}
discusses the implications of $m \neq m_0$, namely that $K$ is a function of
time. This discussion is particularly useful because it does not require
knowledge of the true braking index, $n_0$, which may be different from the
measured value and cannot be predicted from theory. However, since we do not
make any assumption about $n_0$, we cannot say whether $K$ is changing at
the present time. A measurement of $m \neq m_0$ does give us insight into
the variation of $K$ in the pulsar's past. In particular, a value of $m<m_0$ 
implies a period of magnetic field growth early in the pulsar's
history and a true age larger than the characteristic age, while $m>m_0$
reduces the inferred age of the pulsar.  Given
$m-m_0=5.04\pm2.99$, and using
Blandford's prescription for the inferred age of the pulsar, 
the true age of \psr\ is 1000$\pm$700\,yr,
just over half of the characteristic
age, $\tau_c = 1691$\,yr. 
\par
The supernova remnant (SNR) believed to be associated
with \psr\ was originally estimated to have an age of $6-21$\,kyr \citep{shmc83}, 
indicating that the 
pulsar's characteristic age was underestimating the age of the system by a 
factor of $\sim3.5-12$. However, 
\citet{gbm+99} showed that the
SNR may be as young as the characteristic age of the pulsar. They
show that the SNR and pulsar are interacting and explain the large size and
unusual morphology of the remnant by hypothesizing that the remnant is
due to an explosion of high kinetic energy or low ejected mass occurring 
on the edge of a cavity, thus accelerating the expansion of the 
remnant. At first glance, the measured value of $m$ appears to worsen the
possible age discrepancy between the pulsar and the SNR. However, the age of
the remnant is clearly not well determined and could be as low as 1000\,yr.

\par
\citet{br88} discuss the implications of a 
constant observed braking index, $n$, where in this case we must make some
\textit{a priori} assumption of the true braking index, typically assumed to
be $n_0=3$. They give several reasons why $n<3$ may 
be observed. The torque may scale with frequency differently, either due
to the outflow of plasma removing a significant amount of angular momentum,
or a non-dipolar magnetic field.  Another interpretation of $n<3$ is that the magnetic
moment is
currently varying in time, that is, $K=K(t)$. If this is the case, we can then
determine a time scale
for torque evolution and a check on the functional form of evolution 
by finding the first two dimensionless derivatives of
$\dot{\nu} = -K(t){\nu}^n$:
\begin{equation}
d_1 = n - n_0 = {\frac{\dot{K}}{K}}\,{ \frac{\nu}{{\dot{\nu}}}},
\end{equation} and

\begin{equation}
d_2 = m - n_0[2n_0 - 1 + 3(n-n_0)] = \frac{\ddot{K}}{K} \frac{{\nu}^2}{{\dot{\nu}}^2}.
\end{equation}
Equation (9) gives a time scale for torque evolution while equation (10)
provides a check for an assumed functional form for $K(t)$. For \psr, 
$d_1=-0.161 \pm 0.003$ and $d_2=3.5 \pm 3$. For exponential growth of
 $K$, this measurement of $d_1$ implies a time scale for growth
of $\sim 20~000$\,yr. If the form of $K$ is exponential we would expect 
that $d_2=(d_1)^2=0.026$, which cannot be ruled out with the present
uncertainties for $d_2$. We find that the present measurements are
similarily in agreement with a power-law form for $K(t)$. The measurement
of $m$ and the large uncertainties, however, are consistent with the
simplest explanation of a constant value of $K$ and thus a constant magnetic
field. 
\par
Despite scatter due to timing noise, the braking index
is constant over 21.3\,yr to within 1.5\%. This result is similar to
that of the Crab pulsar's braking index, reported to be constant to
within 0.5\%, where measurements were obtained between glitch events
and are typically based on five years of data \citep{lps93}. This reinforces
the need for long-term timing to measure the true value of braking indices
to such precision. 

\subsection{Timing noise analysis}
\label{sec:noise_discussion}
\subsubsection{Glitches}
Our phase-coherent analysis shows that, due to the absence of any sudden
discontinuities in the timing residuals,  \psr\ has not glitched in 21.3\,yr.
\psr\ is the only young pulsar known that has never glitched over 
a long period of time.  \citet{lyn99} showed that there is a
correlation not only between $\dot{P}$ and timing
noise, but $\dot{P}$ and glitch activity, although some of the
youngest pulsars (i.e. the Crab) have glitch activity parameters smaller
than expected from the correlation. The explanation for this is that the internal
temperature of the youngest pulsars is too high for large scale
vortex pinning and unpinning. If this is the case, then the fact that
\psr\ has the lowest glitch rate among young pulsars would suggest
that it has the highest internal temperature, in spite its
characteristic age being larger than that of the Crab pulsar.
On the other hand, so-called `anomalous X-ray pulsars' (AXPs) have been
observed to glitch with parameters that are not very different from those
seen in radio pulsars \citep{klc00,kg03,dos+03}.  Given their estimated
surface temperatures and luminosities (measured via X-ray spectral
observations) AXPs appear to be much hotter than any known radio pulsar.
If AXP glitches indeed have the same physical origin as do glitches in
rotation-powered pulsars, this argues that the glitch phenomenon is not
strongly affected by neutron-star temperature.

\subsubsection{Spectral Analysis}
The spectral index for the timing noise of \psr, $\alpha=-4.6\pm1.0$ is
marginally steeper than
that measured for other pulsars. The low-resolution nature of the Deeter 
method, however, provides only a rough estimate of the true 
spectral index of the noise.
The spectral index of the timing noise for the Crab pulsar has been
determined several times, the most recent of which found $\alpha=-3$ as
well as a periodic component with period of $568\pm10$\,days \citep{sfw03}, 
using a modified Fourier analysis (possible in this case due to the 
dense sampling of the Crab timing data from 1982-1989). The spectral index was
 then  verified with the Deeter polynomial method, though this analysis 
does not probe the lowest frequencies (i.e. longest
timescales) available. Previous analyses of Crab timing data 
were consistent with $\alpha \sim -4$ \citep{dee81, cor80}. \citet{ch80}
show that the timing noise of 11 pulsars is consistent with a pure random
walk in phase, frequency or frequency derivative. However, \citet{dmh+95}
show in a study of 45 pulsars that only 5 are consistent with a pure random
walk, while most pulsars exhibit timing noise either due to a 
combination of a random walk and discrete jumps, or inconsistent with either
hypothesis.  Other pulsars have been shown to have spectral 
indices of timing noise
ranging from $\alpha \sim -0.5$ to $-2.4$ \citep{babd99}. Thus 
there is a range of possible spectral indices that
describe timing noise in different pulsars.

\par
Although there is no physical reason why the spectral index must be an even
integer, these are the only cases for which analytical solutions exist. For 
\psr, the measured spectral index is consistent with having a component of
frequency noise, that is, a random walk in pulse frequency corresponding to
$\alpha= -4$. The measured value of 
the spectral index indicates that there may also be a
component due to a random walk in frequency derivative, characterized by
$\alpha = -6$. Although the measured spectral index alone would seem to
indicate a random walk in frequency, the fact that the residuals are 
not Gaussian distributed after fitting for
higher derivatives indicates that this is not the case. 
A random walk in the $k$th derivative 
of phase is mathematically equivalent to a white noise process in the
$(k+1)$ derivative \citep{gro75a}. This should correspond to `white' 
residuals after fitting out the $(k+1)$ derivative, not seen in
the timing residuals of \psr. 

\par
One of the problems in quantifying timing noise is the lack of a
comprehensive theory of its physical causes. In fact
it is possible that there are several causes of timing noise active in 
one object, thus complicating the interpretation of the measured spectral
index. Free precession is an interesting solution to the
timing noise problem because it can be detected from the changes in the
pulse profile as well as a periodicity in the timing residuals \citep{sls00}.
Unfortunately, the data quality and lack of a long baseline for many
pulsars prohibits the discovery of this process. If
undetected, precession would contribute to the apparent timing noise of the
pulsar and the spectral index of the noise, despite the fact that it
is not a random process.  Additionally, other physical causes of noise,
(e.g. magnetospheric torques or random vortex pinning and
unpinning) will contribute to the measured spectral index. 

Another common problem in assessing the cause of timing noise is that
quasi-periodicities are apparent in the timing
residuals of most pulsars. An
artificial quasi-periodicity will change with the length of the data set, while a real
quasi-periodicty will not \citep{klw+99}. To begin to truly 
understand timing noise in pulsars, it is crucial that
new high-resolution spectral methods, capable of discerning discrete
features due to periodicities, are developed. New methods must be able to deal
with both the spectral leakage and data sampling problems that presently
constrain our measurements and thus understanding of timing noise. Another
problem in developing high-resolution methods for very long period noise is
the extremely long time baselines required to observe many periods of the
noise; the only way to solve this is by brute force, that is, by timing pulsars
for hundreds of years. A 
better understanding, both mathematically and physically, of timing noise will
allow for a better understanding of the deterministic spin-down of
pulsars underlying the noise processes.

\acknowledgments
The authors gratefully acknowledge A.Atoyan, S.M.Ransom and M.S.E. Roberts
for helpful discussions on topics relating to this research.The Molonglo
Radio Observatory is operated by the University of Sydney. The Parkes radio 
telescope is part of the Australia Telescope which is funded by the 
Commonwealth of Australia for operation as a National Facility managed 
by CSIRO. This research made use of data obtained from the High
Energy Astrophysics Science Archive Research Center Online Service, provided
by the NASA-Goddard Space Flight Center. VMK is a Canada Research Chair 
and an NSERC Steacie Fellow. Funding for
this work was provided by NSERC Discovery Grant Rgpin 228738-03 and
Steacie Supplement Smfsu 268264-03.  Additional funding came from Fonds de
recherche de la nature et des technologies du Quebec (NATEQ), the
Canadian Institute for Advanced Research, and the Canada Foundation
for Innovation.

\clearpage
  
\normalsize
\begin{figure}
\plotone{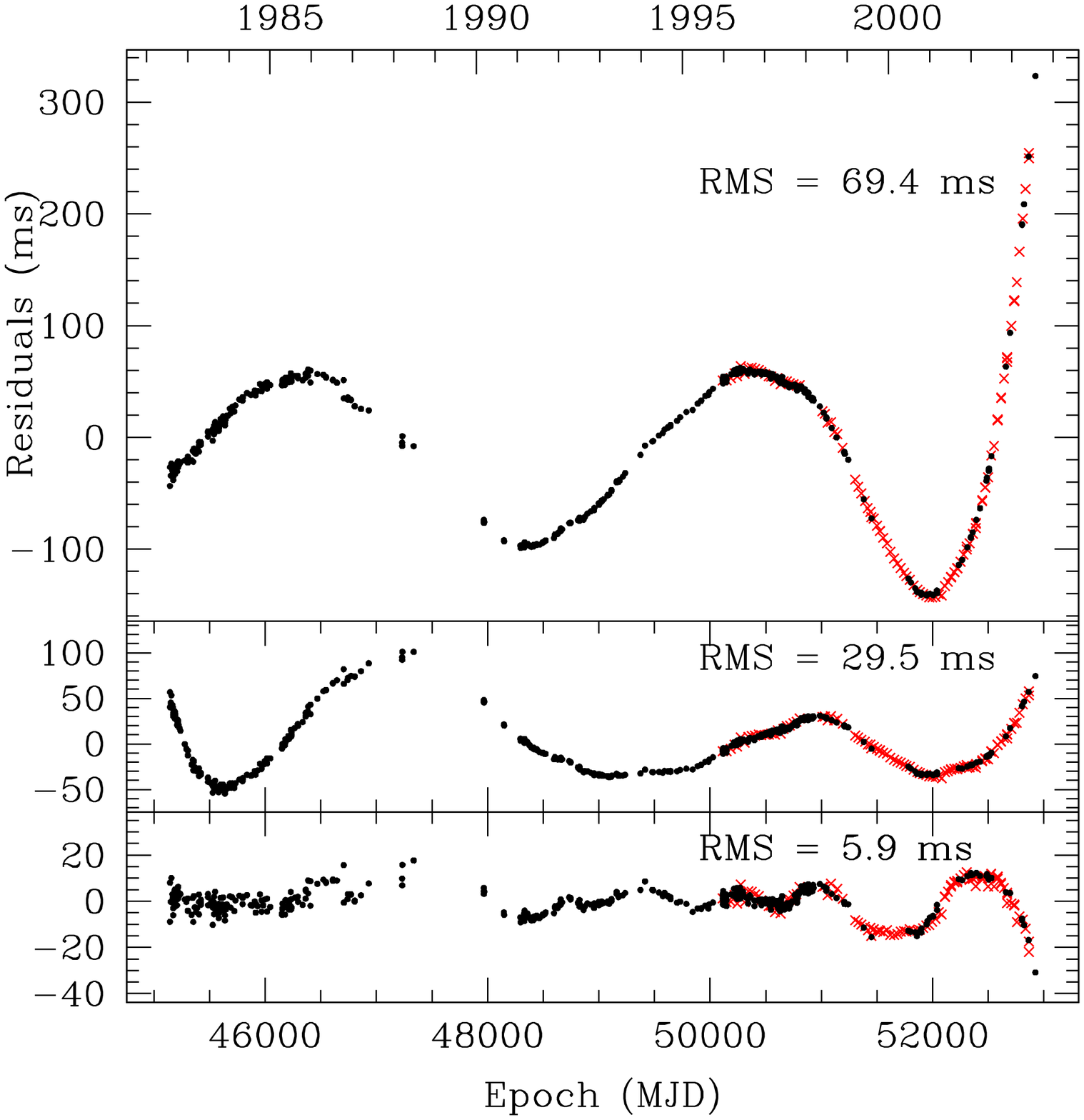}
\figcaption[f1.eps]{Timing residuals for \psr. Radio TOAs are shown as dots; the 
X-ray TOAs as crosses. The top panel has a quartic polynomial (i.e.
$\nudotdotdot$), removed, the middle panel has a quintic removed and the bottom panel
shows the residuals after the removal of a sixth degree polynomial. See the
electronic edition of the Journal for a colour version of this figure. 
\label{fig:residuals}}
\end{figure}

\notetoeditor{In the online version, please include the color version of this 
plot called f1_color.eps}

\normalsize
\clearpage
\begin{figure}
\plotone{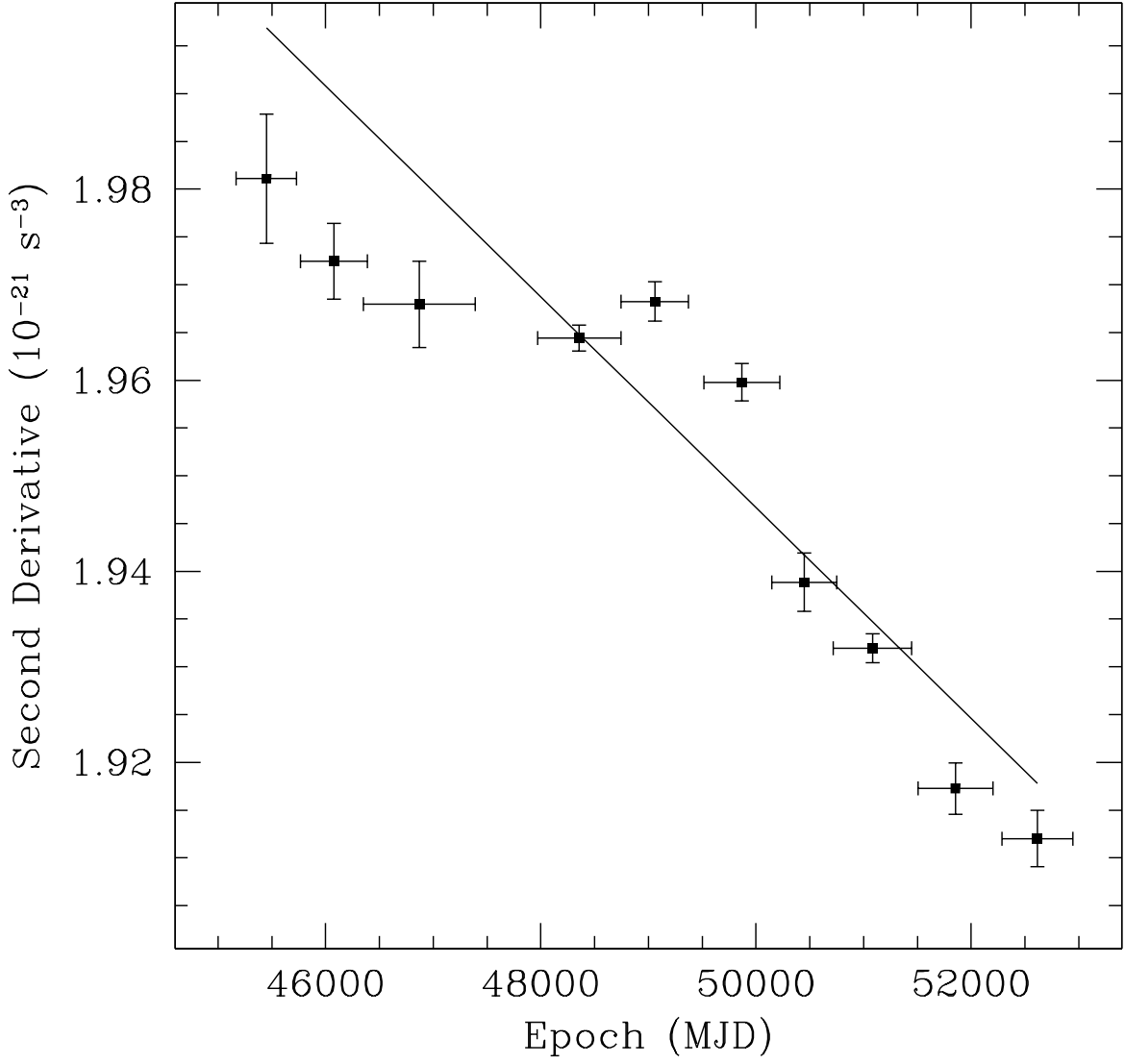}
\figcaption[f2.eps]{Second derivative of pulse frequency from phase-coherent
subsets versus epoch. The slope of the line is $\nudotdotdot$ and 
has a value of $(-1.28 \pm 0.21) \times 10 ^{-31}$\,s$^{-4}$.
\label{fig:f2}}
\end{figure}

\normalsize
\clearpage
\begin{figure}
\plotone{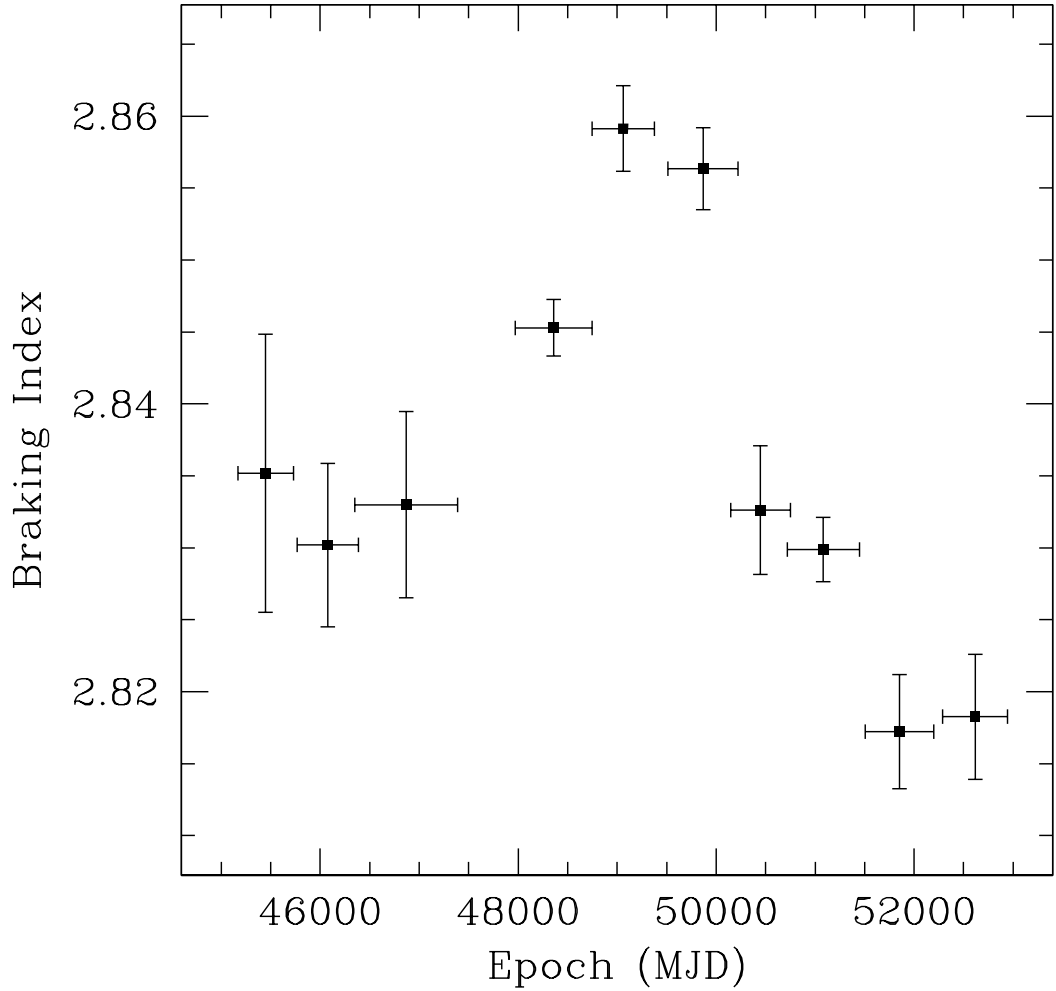}
\figcaption[f3.eps]{Braking index calculated at each epoch. There
is no statistically significant change over 21.3 years of data. 
The average value is $2.839\pm0.003 $, in agreement with the previously
reported value \citep{kms+94} and the value obtained in a fully
phase-coherent analysis (Table 1). The reduced ${\chi}^2$ value is 15 for 9
degrees of freedom, suggesting contamination by timing noise (see Section 3.2).
\label{fig:bind}}
\end{figure}

\normalsize
\clearpage
\begin{figure}
\plotone{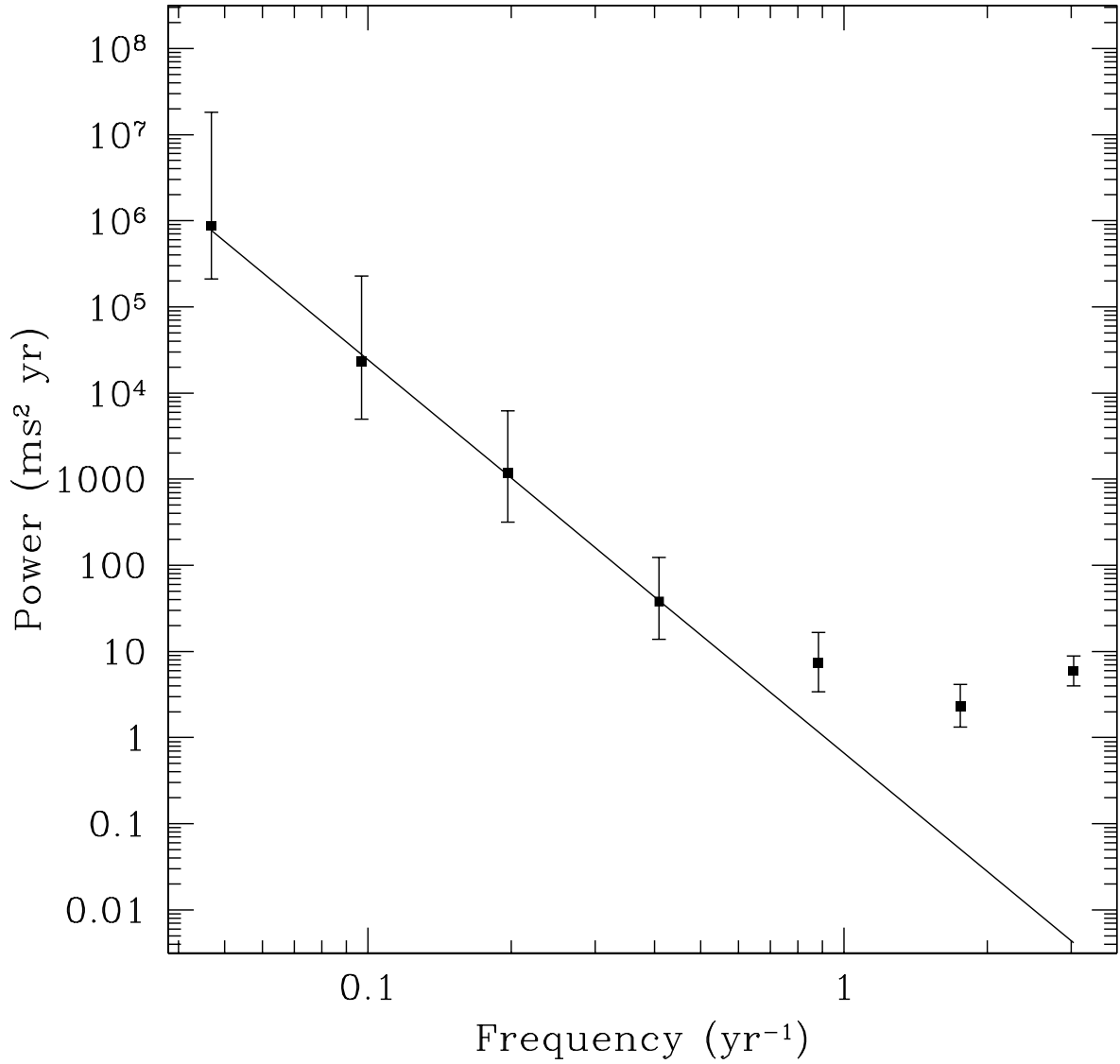}
\figcaption[f4.ps]{Power spectrum of the timing noise exhibited 
by \psr. The best-fit line gives a spectral index of $\alpha = -4.6\pm 1.0$.
The last three points are dominated by measurement errors, thus are not fit as
part of the red noise spectrum.
\label{fig:powerspectrum} }
\end{figure}

\clearpage
\begin{deluxetable}{lc}
\tablecaption{Parameters for \psr.
\label{spin}}
\tablewidth{0pt}
\startdata
\cutinhead{Parameters for phase-coherent analysis.}
Dates (Modified Julian Day)         &  45114 - 52925 \\
Epoch (Modified Julian Day)         & 49034.5    \\
Right Ascension \tablenotemark{a} (J2000)  &15h~13m~55.62s\\
Declination  \tablenotemark{a} (J2000)     &$-59^{\circ} 08$$'$$09.0$$''$ \\
$\nu$ (Hz)                          & 6.633598804(3)    \\
$\dot{\nu}$ ($10^{-11}$~s$^{-2}$)   & $-$6.75801754(4) \\
$\ddot{\nu}$ ($10^{-21}$~s$^{-3}$)  &  1.95671(2)\\
Braking Index, n                       & $2.84209(3)$ \\
Dispersion Measure \tablenotemark{a} (pc~cm$^{-3}$) & 253.2 \\
\cutinhead{Parameters for partially coherent analysis.}
Braking Index, n                     &  2.839(3) \\
$\nudotdotdot$ ($10^{-31}$~s$^{-4}$) &  $-1.28(21)$ \\
Second braking index, m             & $18.3(2.9)$ \\
\enddata
\tablenotetext{a}{Held fixed for phase-coherent analysis.}
\end{deluxetable}

\begin{deluxetable}{lc}
\tablecaption{Dispersion Measure at four epochs.\label{dm}}
\tablewidth{0pt}
\tablehead{
\colhead{Epoch} & \colhead{DM (pc~cm$^{-3}$)} \\}
\startdata
March 15 1990                &  253.2$\pm$1.9  \\
February 1 - February 4 1996 &  257.4$\pm$1.2  \\
February 29 - March 1 1996   &  255.5$\pm$0.9 \\
January 12 - January 14 1997 &  254.8$\pm$1.1 \\
\enddata
\end{deluxetable}

\end{document}